\documentclass[bibyear]{aa} 

\usepackage{pdflscape}
\usepackage{txfonts}
\usepackage{graphicx}

\usepackage{natbib}
\usepackage{gensymb}

\bibpunct{(}{)}{;}{a}{}{,}
\usepackage[colorlinks=true, linkcolor=blue, citecolor=blue, urlcolor=blue]{hyperref}

\newcommand{\halpha}{H\ensuremath{_{\alpha}}}
\newcommand\kms{\mbox{\,km~s$^{-1}$}}

\titlerunning{Discovery of an ultra-thin filament from deep imaging with amateur telescopes}
\authorrunning{D. Martínez-Delgado et al.}
\usepackage{natbib}
\bibpunct{(}{)}{;}{a}{}{,} 
\usepackage{gensymb}

\usepackage{pdflscape}
\usepackage{longtable}

\usepackage{graphicx}
\usepackage{txfonts}

\begin{document}

\title{Discovery of an extremely narrow trail-like feature crossing the Veil Supernova Remnant in deep amateur observations}

   \author{David~Mart\'inez-Delgado\inst{1,2}\fnmsep\thanks{ARAID Fellow},  Jan-Niklas Pippert \inst{3,4}, Miguel A. P\'erez-Torres\inst{5,9,10}, Malte Busmann \inst{4,11}, Emilio J. Alfaro\inst{5},  Mark Hanson\inst{6}, Giuseppe Donatiello\inst{7}, Alexander Zaytsev\inst{6}, Claudia Toci\inst{8},  Claus Gössl \inst{4}, Michael Schmidt \inst{4}, Christoph Ries \inst{4}}
\institute{
$^{1}$ Centro de Estudios de F\'isica del Cosmos de Arag\'on (CEFCA), Unidad Asociada al CSIC, Plaza San Juan 1, 44001 Teruel, Spain\\
$^{2}$ ARAID Foundation, Avda. de Ranillas, 1-D, E-50018 Zaragoza, Spain\\
$^{3}$ Max Planck Institute for Extraterrestrial Physics, Giessenbachstr. 1, D-85748 Garching, Germany \\
$^{4}$ University Observatory, Faculty of Physics, Ludwig-Maximilians-Universität München, Scheinerstr. 1, 81679 Munich, Germany \\
$^{5}$ Instituto de Astrofísica de Andalucía (CSIC), Glorieta de la Astronom\'\i a,  E-18080 Granada, Spain\\
$^{6}$Doc Greiner Research Observatory-Rancho Hidalgo, Animas, New Mexico, USA \\
$^{7}$UAI -- Unione Astrofili Italiani /P.I. Sezione Nazionale di Ricerca Profondo Cielo, 72024 Oria, Italy \\
$^{8}$ Departamento de Fisica Aplicada III, Universidad de Sevilla, E.T.S.I.., Camino de los Descubrimientos, 41092 Sevilla, Spain\\
$^{9}$ School of Sciences, European University Cyprus, Diogenes street, Engomi, 1516 Nicosia, Cyprus\\
$^{10}$ Depto. de F\'isica Te\'orica, Universidad de Zaragoza, E-50009 Zaragoza, Spain \\
$^{11}$ Excellence Cluster ORIGINS, Boltzmannstr. 2, 85748 Garching, Germany \\
}

   \date{Received X; accepted Y}

 
  \abstract
  {}
   {We report the discovery of an extremely narrow and highly collimated gaseous trail-like structure crossing the eastern region of the Veil Nebula (Cygnus Loop). This study aims to characterize the morphology and photometric properties of this previously unknown feature and to investigate its physical origin}
   {The feature was first identified in deep narrow-band \halpha images obtained by a small telescope and was confirmed in more than a dozen publicly available amateur and professional images in the last two decades, rejecting the possibility of being an artifact (e.g., artificial satellite trail). Follow-up observations were performed with the 2.1 m Fraunhofer Telescope Wendelstein using 
   \halpha, [SII], and $i$-band filters. To characterize its structural and luminosity parameters, we applied a model-fitting code originally developed for the analysis of extragalactic stellar tidal streams.}
   {The structure is only detected in \halpha emission, with no detectable counterpart in [SII] or in visible observations, and shows an almost constant brightness and width along its extension. It exhibits a surface brightness of 22.32 $\pm$0.13 \halpha mag \arcsec$^{2}$. Model fitting yields a median width of 1.63\arcsec, which corresponds to a physical scale of approximately 600 UA, assuming it is at the same distance (2400 light-years) as the Veil Nebula remnant.}   
   {We discuss several potential scenarios for the origin of this feature: a Herbig-Haro-like jet, the trail of a high-velocity object, or a non-radiative shock associated with the supernova remnant. 
   We rule out the Herbig-Haro scenario, given the absence of a driving stellar source and the lack of [S II] emission. A high-velocity-object wake cannot yet be excluded, particularly if the driver is a compact remnant, but the absence of an obvious source or bow-shock apex and the extremely small, nearly constant width of the structure make a normal runaway-star origin unlikely. The current evidence instead favours a Balmer-dominated non-radiative shock associated with the Cygnus Loop, generated as the SNR blast wave encounters a dense material layer or magnetic structure viewed nearly edge-on.
   Further deep imaging and spectroscopic observations with larger telescopes are needed to distinguish between these scenarios.}

   \keywords{ISM: individual objects: Veil Nebula (Cygnus Loop) – ISM: supernova remnants - Shock waves}

   \maketitle
    \nolinenumbers
   
%

\section{INTRODUCTION}

\begin{figure*}
\begin{center}
\includegraphics[width=1\textwidth]
{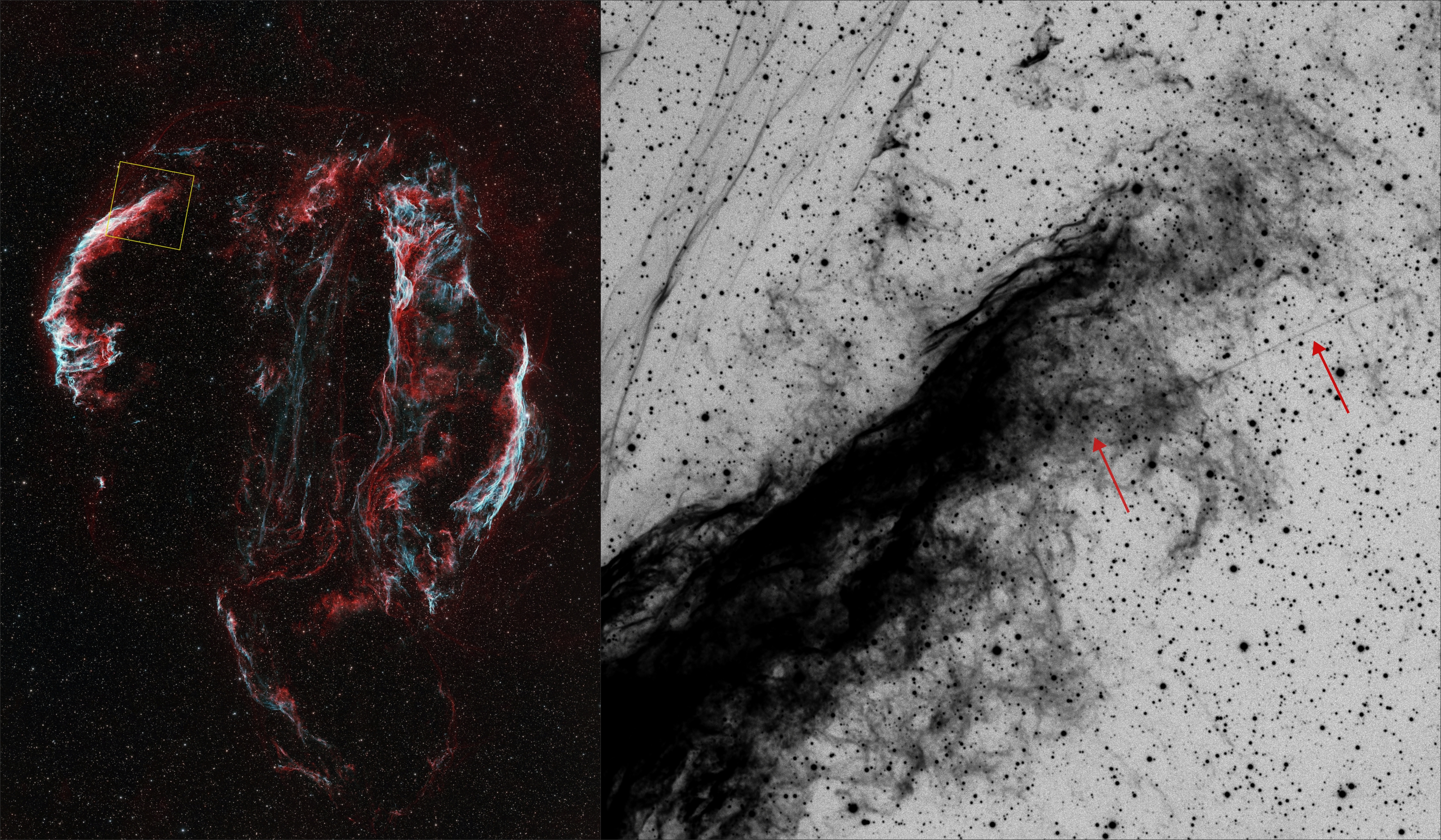}
\caption{Wide-field image of the Vela Supernova Remnant ({\it left}) and deep H$\alpha$ image of the region analysed in this work ({\it right}). The box in the left panel marks the field of view of the H$\alpha$ image. The wide-field image is a $\sim$6 h RGB and narrow-band stack obtained at Piano Visitone (Pollino National Park, Italy) by Giuseppe Donatiello with Tair 3-S 300 mm $f/4.5$ telephoto lenses and a custom 110/250 mm $f/2.2$ astrograph equipped with DSLR cameras and an Optolong 3 nm Dual-Band L-Ultimate filter. The H$\alpha$ image was obtained by Alexander Zaytsev and Mark Hanson at the Dark Sky New Mexico Observatory (Stan Watson Observatory) with a 43 cm Planewave CDK telescope ($f/6.8$) and an SBIG STX-16803 CCD camera, with a total exposure time of 10.5 h. The trail feature discussed in this paper is identified with an arrow.}
\label{fig:GTC}
\end{center}
\end{figure*}

The Veil Nebula is one of the oldest and most extensively studied supernova remnants (SNRs) in the Milky Way. It originated from a core-collapse supernova explosion that occurred approximately 8,000–10,000 years ago, leaving behind an expanding shell of ionized gas and shock fronts that currently span several degrees projected into the sky in the constellation Cygnus. Its proximity (at a distance of approximately 0.73 kpc; Fesen et al.~\citeyear{Fesen2021}), large angular size ($\sim$ 3$\degree$ in total; see Fig \ref{fig:GTC}; left panel), and rich filamentary structure have made it a benchmark target for understanding the interaction between supernova ejecta and the interstellar medium (ISM), as well as the long-term evolution of shock-heated plasmas \citep{Blair2005}. 

The complex morphology of the Cygnus Loop is primarily shaped by the interaction between the expanding blast wave and interstellar clouds spanning a wide range of densities \citep{Fesen2021CygnusLoop}.  Among the resulting features, Balmer-dominated filaments are of particular interest because they trace non-radiative collisionless shocks propagating through partially neutral gas, producing H$\alpha$ emission through collisional excitation and charge exchange, often with broad and narrow Balmer components but weak forbidden-line emission \citep{Hester1986,Hester1994,Ghavamian2001,Heng2010,Raymond2023}.
Although large-scale shock fronts of the remnant have been extensively studied in the literature, the identification of very highly collimated and narrow filamentary features offers a valuable opportunity to probe localized density enhancements, magnetic-field structures, or the wakes generated by high-velocity objects \citep{CantoRaga1998, Raga2000}.  The detection and characterization of these small-scale phenomena are essential for improving our understanding of collisionless shock dynamics and the mechanisms responsible for particle acceleration in supernova remnants \citep{Vucetic2023}.

 In addition to its scientific interest, the Veil Nebula is also one of the most frequently imaged targets by astrophotographers, due to its striking morphology and strong emission-line contrast. In recent years, advances in amateur equipment and commercial detector sensitivity have yielded deep images reaching extraordinarily low surface-brightness levels, revealing previously undetected diffuse structures surrounding the well-known main filaments. These observations have allowed us to probe the outermost regions of the remnant with unprecedented depth, opening a new parameter space for studying its full dynamical and morphological complexity.  In this Letter, we report the discovery of an unusually thin, highly collimated trail-like structure crossing the eastern region of the Veil Nebula (NGC 6992). The feature is visible only in deep H$\alpha$ observations independently obtained by multiple amateur telescopes and was already detected (but not recognised) in \cite{Vucetic2023}, highlighting both its extremely low surface brightness and the remarkable sensitivity now achievable with modern amateur instrumentation. Motivated by its unusual morphology, we obtained high-resolution follow-up observations with the 2.1 m Fraunhofer Telescope Wendelstein. These data allow us to characterize the structure in detail and to assess possible scenarios for its origin.

\section{OBSERVATIONS}
\label{sec:meth}

\subsection{Amateur images}

The narrow feature was first identified in deep images of the Veil Nebula obtained at the Dark Sky New Mexico Observatory (Stan Watson Observatory; see right panel in Fig \ref{fig:GTC}) with a remotely operated 43 cm Planewave CDK telescope ($f/6.8$). The observations were acquired with an SBIG STX-16803 CCD camera, providing a pixel scale of 0.63\arcsec,pix$^{-1}$ and a field of view of $43 \times 43$\arcmin. A total of 22 exposures of 1800 s were obtained in H$\alpha$ ($\lambda = 656.3$ nm) and nine exposures of 1800 s in [S,II] ($\lambda = 671.6$ nm) using Astrodon 3 nm narrow-band filters. The data were collected during several photometric nights between August and October 2018. Individual frames were processed following standard CCD reduction procedures, including bias subtraction, dark-current correction, and flat-fielding. The reduced images were then aligned and combined to produce final co-added images with total integration times of 660 min in H$\alpha$ and 270 min in [S,II], corresponding to a cumulative exposure time of 930 min.

\subsection{Fraunhofer Wendelstein Telescope}
\begin{figure*}
    \centering
    \includegraphics[width=\linewidth]{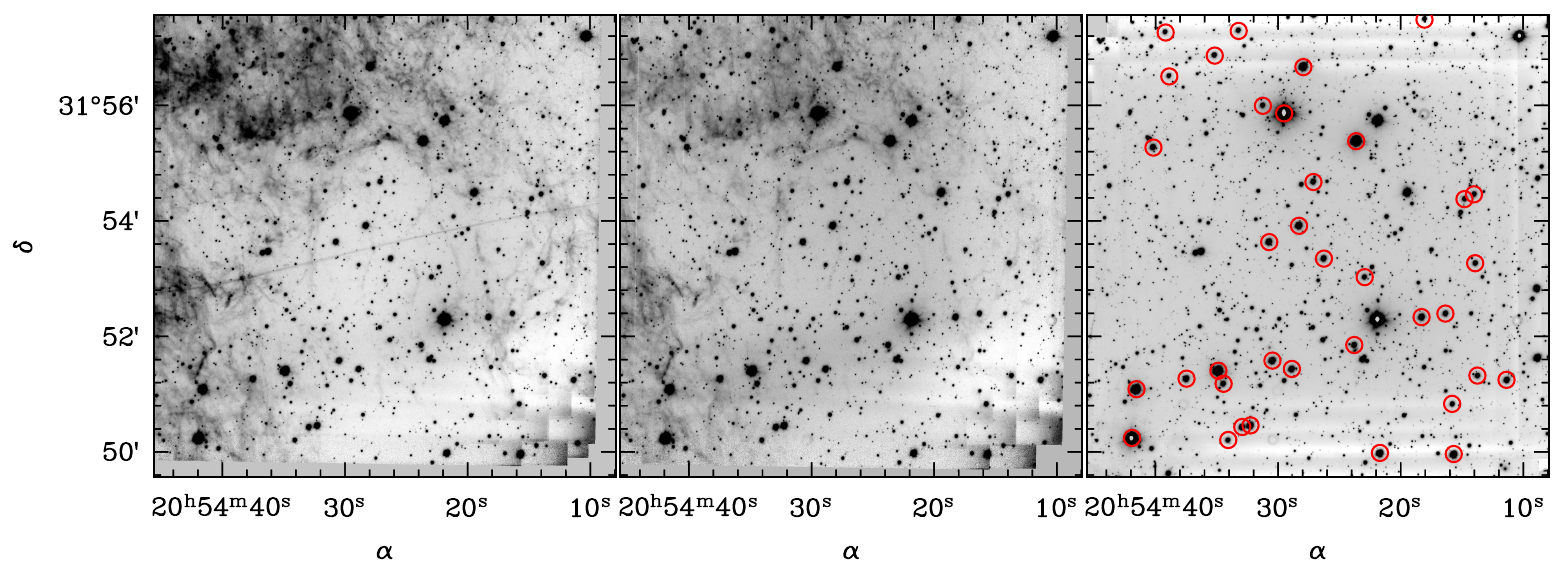}
    \caption{Final coadds of the observations with the 3kk camera. \textit{(left)} $H_\alpha$. \textit{(middle)} $SII$. \textit{(right) $i$}. Red circles indicate the Gaia DR3 spectra sources used for the zero-point calibration. For $H\alpha$ and [SII] we checked that the saturated $i$ band stars behave linearly.}
    \label{fig:3kk_stacks}
\end{figure*}

Scientific follow-up observations were carried out with the $2.1\,$m Fraunhofer Telescope Wendelstein (FTW), from 30th of October 2025 until 20th of January 2026, operated by the University Observatory of the Ludwig Maximilians University of Munich (Fig.\ref{fig:3kk_stacks}). The observatory is located on Mount Wendelstein, approximately $70\,$km south of Munich, at an altitude of $1838\,$m. The observations were obtained with the 3KK camera, a three-channel imaging instrument with a pixel scale of $0.2$ arcsec pixel$^{-1}$ and a field-of-view of $7'\times7'$. Standard CCD reduction procedures were applied to all images, including bias subtraction, flat-field correction, cosmic-ray removal, and astrometric calibration. The pipeline also determines the photometric zero point; however, as observations were carried out in the narrow bands $H\alpha$ and [SII], the zero point was determined using the approach described in Appendix \ref{app:zeropoints}.

\section{ANALYSIS AND MODEL FITTING}

The trail resembles a straight, very collimated stream crossing the Veil supernova remnant, making it particularly well suited for morphological analysis using our code designed for the analysis of extragalactic stellar tidal streams in nearby galactic halos described in \cite{Pippert_2025}. A major limitation, however, is the strong contamination from the remnant emission itself. Nevertheless, we have selected a localized region with minimal nebulosity clouds, providing suitable conditions for a reliable fit to characterize the luminosity and structural parameters of this feature. Isolating the emission from the background procedure is analogous to what is routinely done to perform analysis of these structures (see \citealp{Blair2005, Vucetic2023}).

\begin{figure}
    \centering
    \includegraphics[width=\linewidth]{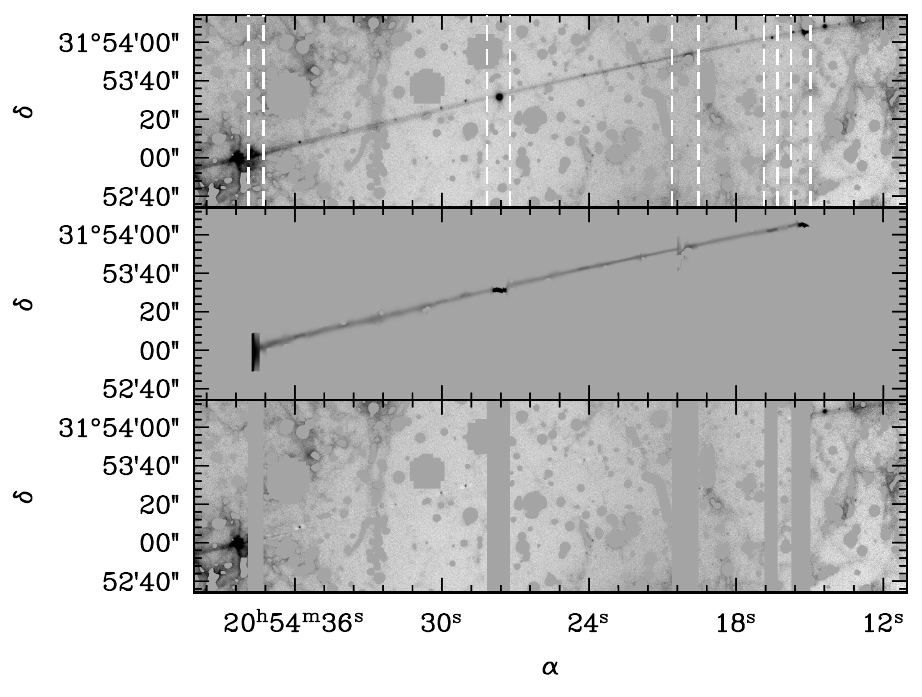}
    \caption{Summary of the stream fitting. \textit{(top)} The masked cutout of the input image. Vertical white dashed lines mark areas that are excluded after the fit for the analysis, due to highly contaminated pixels lying directly on the stream. \textit{(middle)} The 2D model using \texttt{astrostreampy}. \textit{(bottom)} The masked residual image. Here, the marked stripes from the top panel are filled with zero pixels.}
    \label{fig:streampy_halpha}
\end{figure}

The fitting procedure is performed by dividing the feature into a series of boxes, with each box corresponding to a small cutout whose width and height are adapted to the size, morphology, and signal-to-noise ratio (S/N) of the trail feature. An initial fit is first obtained within a high-S/N region of the feature. At these coordinates, the central one-dimensional pixel slice of the box, taken perpendicular to the elongation direction, is used as the reference model. The box is then shifted by one pixel along the feature, and the fitting process is repeated iteratively. The model is composed of a spatial grid defined by the parameters $(x_0, y_0, w, h, \theta)$, together with a Gaussian profile characterized by its amplitude, dispersion $\sigma$, Gauss--Hermite moments $(h_2, h_3, h_4)$, and an additive offset. The Gaussian profile is injected into the grid, producing an averaged representation of the feature within each box. A more detailed description of the method is provided in \cite{Pippert_2025}. The results are shown in Fig. \ref{fig:streampy_halpha}.

The model not only inherits the 2D model image but also a description of the feature at each x and y coordinate. That includes the width (sigma of the Gaussian), its amplitude (brightness), and the local background (offset). We produce width and brightness profiles based on the fit parameters. While the model is a Gaussian, we can create photometric apertures governed by, e.g., a 3xFWHM criterion and hence derive the surface brightness of the stream.

Performing photometry on this feature is challenging as the supernova remnant itself highly contaminates the field. This significantly influences the estimation of the local background. We therefore rely on the created model to measure the surface brightness of the stream, since the local background is fitted in each iteration; hence, local contamination is taken into account. We arrive at a surface brightness of $22.32\pm0.13$ $H_\alpha$ mag $\mathrm{arcsec}^{-2}$.

\begin{figure}
    \centering
    \includegraphics[width=\linewidth]{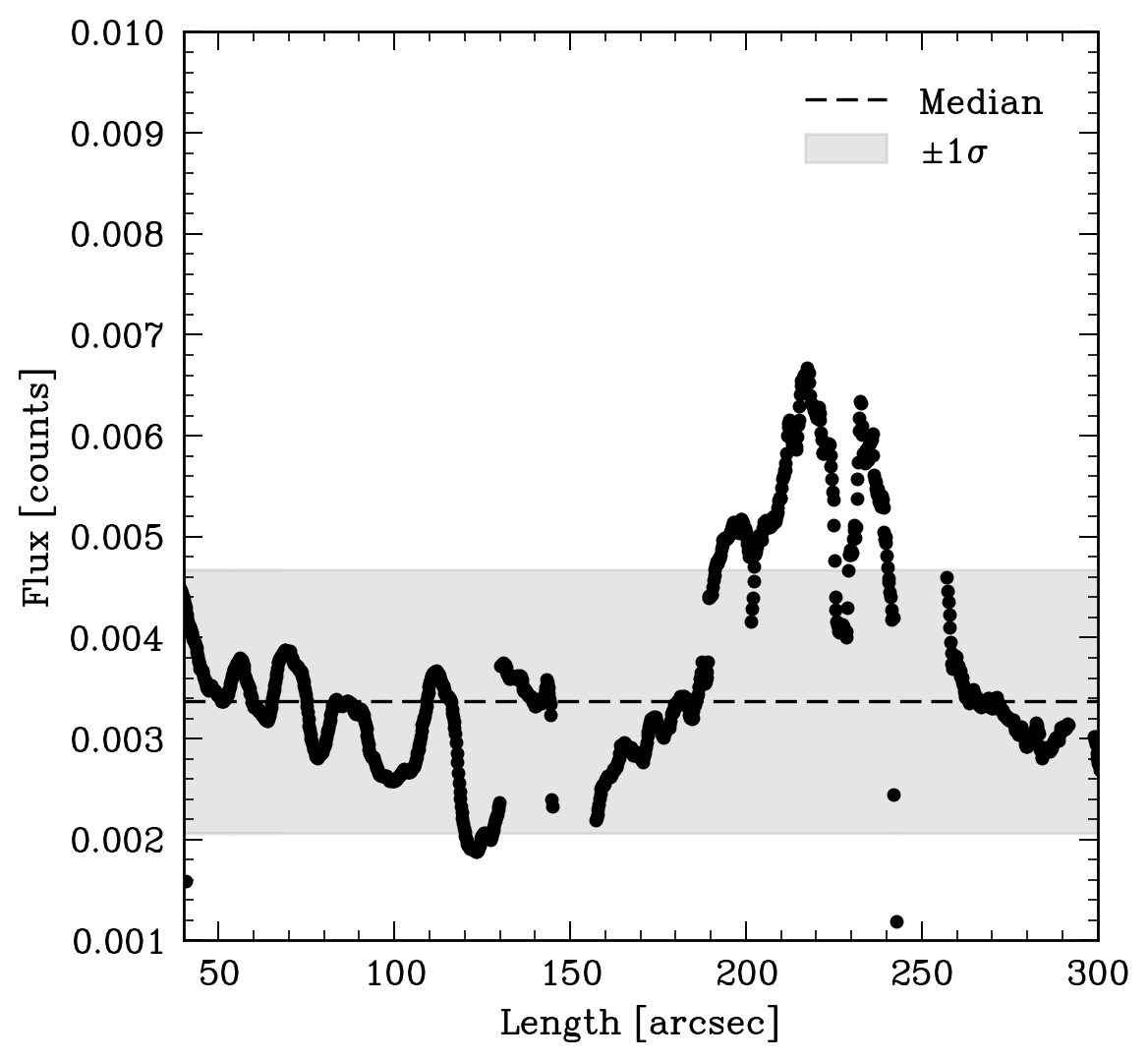}
    \caption{The 1D brightness profile extracted from the best fit amplitude of each Gaussian from the fitting.}
    \label{fig:curvature}
\end{figure}

\begin{figure}
    \centering
    \includegraphics[width=\linewidth]{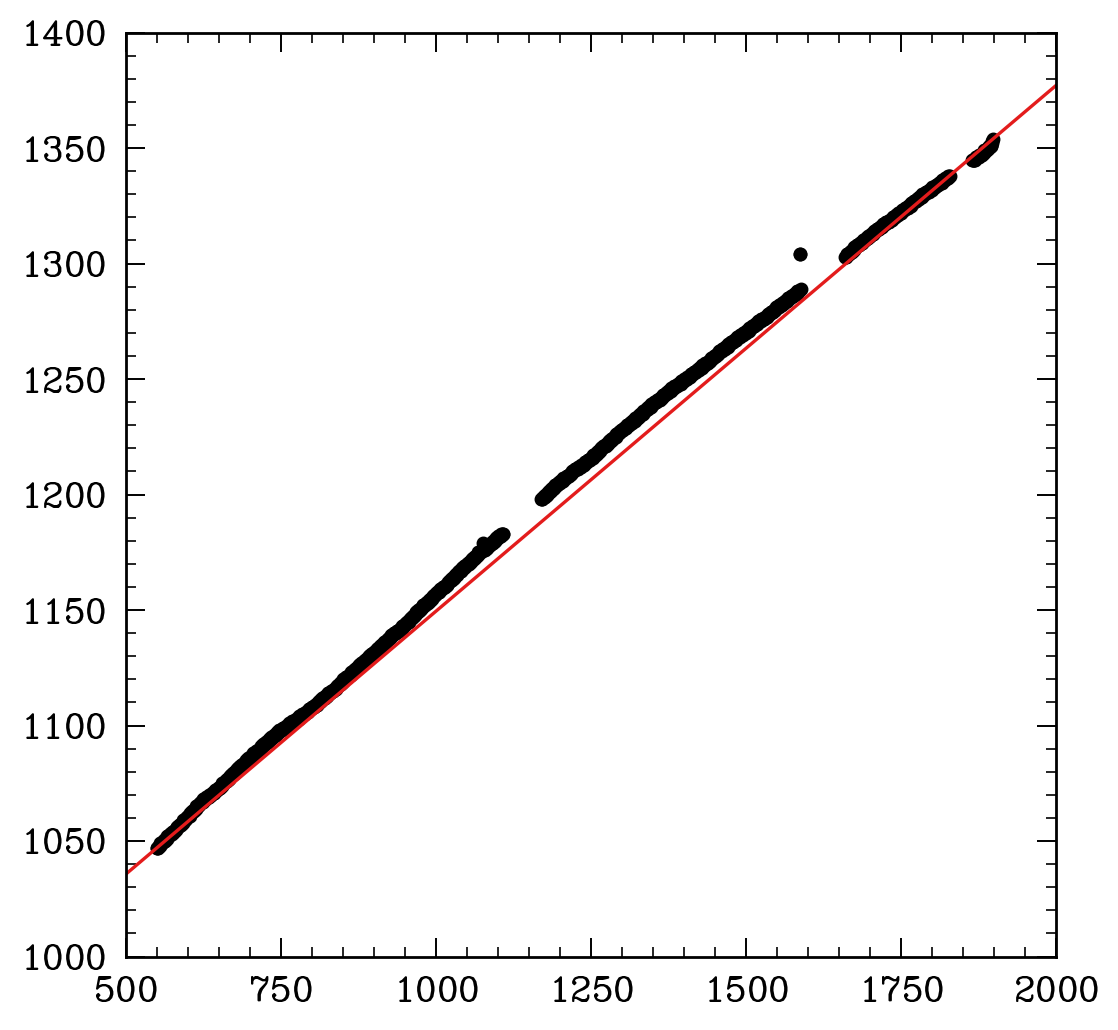}
    \caption{Curvature of the stream. The black points are the box centers of the fitting. Red is a linear relation between two outer points of the fitted segment.}
    \label{fig:placeholder}
\end{figure}

From the fitted parameters, we compute a median width of the stream at one FWHM to be $1.630$ arcsec, which corresponds to a physical scale of $5.8 \times 10^{-3}$ pc, or about 1200 AU, for an assumed distance to the remnant of $\sim0.73$ kpc \citep{Fesen2021}. The 1D width and brightness profiles show a noisy graph, from which we infer a constant brightness and width along the stream (Fig.\ref{fig:curvature}) . Further, the stream is not completely straight. The modeled segment clearly shows a noticeable curvature (Fig.\ref{fig:placeholder}).


\section{DISCUSSION}
\label{sec:disc}

The origin of the linear \halpha\ feature crossing the Veil Nebula is uncertain. Its morphology and narrow-band data are compatible with several distinct physical interpretations, each involving different mechanisms for producing highly collimated ionized structures. In this section, we discuss the most plausible scenarios in light of the available observations, considering their consistency with the absence of detectable  [SII] emission with small amateur telescopes.

\subsection{A Herbig-Haro (HH)-like jet} 

A straight, collimated filament seen in \halpha\ recalls Herbig-Haro (HH) jets. HH objects are shocked outflows from young stellar objects and typically exhibit knotty morphologies and strong low-excitation forbidden lines, especially [S II] and [O I], in addition to H$\alpha$ \citep{Reipurth2001}.
The absence of detectable [S~II] emission, together with the lack of an obvious driving young stellar object at either end of the structure, strongly argues against an interpretation as a Herbig--Haro (HH) jet. To further assess whether the observed emission could arise from an HH shock detected only in H$\alpha$, we estimated the expected proper motion of the emitting knots.

HH jets typically exhibit transverse velocites of the order of tens to several hundreds of km s-1, with H$\alpha$ emission tracing shocked gas, while multi-epoch HST observations have revealed HH ejecta moving at velocities up to $\sim 300\ \mathrm{km\ s^{-1}}$ \citep[e.g.,][]{Hartigan2005}. Using the standard relation
\begin{equation}
v_{\rm tan} = 4.74\,\mu\,d,
\end{equation}
where $v_{\rm tan}$ is expressed in $\mathrm{km\ s^{-1}}$, $\mu$ is the proper motion in $\mathrm{mas\ yr^{-1}}$, and $d$ is the distance in kpc, we obtain an expected displacement of approximately $0.3$--$1.7$ arcsec over a 20 yr baseline, assuming that the jet lies close to the plane of the sky. No such positional shift is detected in our multi-epoch images (spanning 22 years), providing additional evidence against the HH-shock scenario.

However, we cannot fully exclude the HH-jet, as the missing [SII] emission could be a result of missing depth and the inclination of emission can be particularly unlucky. We estimate the $3\sigma$ detection depth in a $10''\times10''$ in the [SII] 3kk stack, following \cite{roman+2020}, to be between $\sim25.21$ and $\sim25.96$ mag arcsec$^{-2}$. This limit depends strongly on how conservative one masks contaminating sources in that highly contaminated field. Compared to low surface brightness studies with broadband imaging, a depth of $\sim26$ mag arcsec$^{-2}$ is rather low. Deeper narrowband observations might reveal faint [SII] emission.




\subsection{A high-velocity object}

A second possibility is that the feature traces the wake of a high-velocity object moving through gas within, or projected against, the Veil Nebula. This class of explanations includes two physically different scenarios: a stellar runaway, in which the trail would be produced by a stellar wind bow shock and/or by photoionization from a hot star, and a compact-object wake, such as a neutron-star or pulsar bow shock propagating through partially neutral gas. In the classical stellar-wind case, the characteristic scale of the bow shock is set by ram-pressure balance,
\[
R_0=\left(\frac{\dot{M}v_{\rm w}}{4\pi \rho_{\rm a}v_*^2}\right)^{1/2},
\]
and the morphology is expected to include an identifiable stellar source, an apex or bow-shaped head, and a downstream wake whose width is controlled by the wind momentum, stellar velocity, and ambient density \citep{Wilkin1996,Peri2012,Peri2015,Meyer2016}. Photoionized trails from hot runaway stars are also expected to be accompanied by a detectable ionizing source and a broader H II/bow-shock structure rather than by an isolated, ultra-thin H$\alpha$ filament 
\citep{Canto1998,Raga2000,Raga2022}.

The present observations make the ordinary hot-runaway-star interpretation unlikely. At the adopted distance of the Cygnus Loop, the measured width corresponds to only a few $10^{-3}$ pc, while the projected length appears to be of order parsecs, implying an extreme aspect ratio. Moreover, the feature shows an approximately constant width and brightness along the fitted segment, lacks an obvious bow-shock apex, and has no detected driving star in the available optical images. The absence of detectable [S II] emission further argues against a radiative stellar bow shock or a Herbig--Haro-like outflow. We therefore regard a normal stellar runaway as a disfavoured explanation, although not formally excluded until deeper imaging and a dedicated search for high-proper-motion sources have been performed.

A compact-object wake is harder to exclude. H$\alpha$ bow shocks are known around high-velocity pulsars moving through partially neutral gas, where Balmer emission traces the interaction between the pulsar wind, the stellar motion, and the neutral ambient medium 
\citep{Chatterjee2002,Brownsberger2014,deVries2020,deVries2022}. Under this interpretation, the feature should terminate near a compact driver, possibly associated with an X-ray or radio counterpart, and the trail should show a coherent velocity and/or surface-brightness gradient along the direction of motion. The absence of such a source in current data would therefore not by itself rule out the compact-object scenario, but it makes it less natural than a shock-front geometry intrinsic to the Cygnus Loop.

However, this scenario is also difficult to reconcile with the available observations. Typical runaway stars cover a broad range of physical scales and velocities, spanning from velocities of $v \sim 30\ \mathrm{km\ s^{-1}}$ \citep{2011MNRAS.410..190T} for massive OB stars to velocities of $v \sim 600\ \mathrm{km\ s^{-1}}$ \citep{2004ApJ...600L..51C} 600 km/s for compact objects such as neutron stars. Considering the available imaging data obtained over the last two decades, and assuming that the observed feature corresponds to the leading tip of the trail, we do not detect any significant positional change. For the expected stellar velocities, the trail tip should have advanced by approximately 130–4000 AU during this period. At the distance of the Veil Nebula, this displacement corresponds to an angular shift of about 0.2$\arcsec$–5.0$\arcsec$. However, no measurable shift is observed in the position of the trail tip. We note, nevertheless, that the faintness and diffuse morphology of these features make their identification and positional measurements uncertain in the available images. Therefore, although we cannot completely rule out this interpretation yet, the runaway-star hypothesis is not favored by the current data and would require the identification of a stellar counterpart associated to the observed emission.

\subsection{A Non-radiative shock}
The linearity of the filament and the fact that the feature appears only in \halpha, and not as a transient artifact, strongly suggest that the explanation is very likely connected to shocks in partially neutral gas. A natural possibility is that it represents a Balmer-dominated filament produced by a
non-radiative shock front of the remnant itself, intercepting a dense sheet and/or a very
flat, well-ordered magnetic flux tube, seen almost exactly edge-on. 

In such shocks, H$\alpha$ is excited by collisions and by charge exchange of neutrals crossing the shock \citep{Ghavamian2001,Sollerman2003,Heng2010, Morlino_2012}, while the forbidden lines ([S II], [O III]) may be very weak or absent, unlike in radiative shocks \citep{Ghavamian2001,Heng2010,Raymond2023}. The extreme collimation would then arise because we are viewing the line of tangency of that front with the sheet; small three-dimensional undulations would project as a very thin, straight filament 
\citep{Hester1986,Hester1994}.
 If the remnant lies at $\sim$0.73 kpc \citep{Fesen2021} and the shock moves at 
speeds of $\sim (240 - 650)$ \kms\ \citep{Vucetic2023}--and possibly smaller since the shock maybe tangential--we should expect proper motions of the order of 
0.06 - 0.15 arcsec\,yr$^-1$ and an H$\alpha$ profile showing a broad component of a few hundreds of km s$^{-1}$ in addition to the narrow one, a hallmark of Balmer-dominated shocks 
\citep{Ghavamian2001,Sollerman2003,Heng2010}.


\section{SUMMARY AND FUTURE WORK}
\label{sec:concl}

We report the discovery of a remarkably straight, highly collimated gaseous trail embedded in the eastern region of the Veil Nebula within the Cygnus Loop, identified through deep imaging obtained with amateur telescopes. This ultra-thin filament is detected exclusively in \halpha, and its presence in multiple independent datasets acquired over different epochs strongly argues against an artificial origin.\footnote{Recent VLA observations of the Cygnus Loop have been presented by \citet{Urosevic2026}. However, the surveyed area does not include the region studied in this work (see their Fig.~1). Therefore, no radio constraints are available for the trail feature, preventing us from using a potential VLA non-detection as supporting evidence for its classification as a non-radiative filament.}

We discuss three possible interpretations for this exceptionally narrow H$\alpha$ filament: (i) an HH-like jet; (ii) the wake of a high-velocity object moving through gas within, or projected onto, the Veil Nebula; and (iii) a Balmer-dominated non-radiative shock associated with the supernova remnant encountering a dense sheet or ordered magnetic structure. The HH-like jet scenario is strongly disfavoured by the lack of a driving young stellar object and by the current non-detection of [S II]. A normal hot runaway star is also less likely, since such systems are expected to show an identifiable stellar driver, a bow-shock apex, and/or a broader photoionized structure. A compact-object wake remains possible, but would require the detection of a high-proper-motion source, an X-r ay or radio counterpart, or a coherent kinematic signature along the filament. In addition, the runaway-object scenario is also disfavored by the lack of any discernible morphological or positional changes between the 1993 image and the more recent observations. Unless the object's velocity vector is oriented predominantly along the line of sight, minimizing its projected motion on the plane of the sky, a detectable displacement would be expected over the available temporal baseline of the available images. On the basis of the current data, the most plausible working interpretation is therefore that the feature is a Balmer-dominated, non-radiative shock related to the Cygnus Loop itself, seen where the blast wave intercepts a thin sheet of material or a magnetic structure nearly edge-on.

Deep narrow-band imaging in H$\alpha$ and [S II] $\lambda\lambda6716,6731$ with a large professional telescope can provide the decisive diagnostic of the physical origin of this linear feature in the Cygnus Loop. The H$\alpha$/[S II] ratio is a well-established tracer of shock conditions
and ionization state in supernova remnants.

In radiative shocks, cooling layers efficiently produce forbidden-line emission, typically yielding $\mathrm{[S,II]}/\mathrm{H}\alpha \gtrsim 0.4$ 
\citep{Fesen1996,Stupar2012}, whereas Balmer-dominated, non-radiative shocks propagating into partially neutral gas are characterized by strong H$\alpha$ emission, broad and narrow Balmer components, and weak or absent forbidden lines 
\citep{Ghavamian2001,Heng2010,Raymond2023}. Therefore, measuring the
H$\alpha$/[S II] ratio, or placing stringent limits on it,
constitutes a key test of the nature of the structure and a
standard diagnostic for distinguishing radiative and
Balmer-dominated shocks \citep{Heng2010,Raymond2023}.
Such a measurement can discriminate between an atypical manifestation of the Cygnus Loop shock, arising from its interaction with a localized density enhancement or an ordered magnetic field structure, and a physically distinct phenomenon, such as the wake of a high-velocity compact object or an unrelated emission feature projected along the line of sight. Together with the striking linear morphology and the absence of detectable variability over independent epochs, a reliable line-ratio determination should provide the quantitative basis needed to establish the physical origin of this feature.

Equally important, a non-detection of [S II] emission in extremely deep observations, confirming that the structure remains detectable only in H$\alpha$,  would favor a Balmer-dominated non-radiative shock propagating through partially neutral material, potentially viewed along an exceptionally thin tangency within the Cygnus Loop shock front. In this case, the feature would provide an unusual probe of collisionless shock physics, ion--neutral interactions, and particle acceleration processes in supernova remnants. The absence of measurable forbidden-line emission would also argue against interpretations involving radiative shocks, Herbig--Haro-like outflows, or other phenomena characterized by cooling post-shock regions. Conversely, the detection of significant [S II] emission would point to the presence of cooling gas and require a substantially different physical interpretation. Deep narrow-band imaging therefore represents a critical observational test capable of discriminating between competing scenarios and clarifying the origin of this enigmatic structure.

\begin{acknowledgements} 

DMD acknowledges all the astrophotographers who generously contributed their images of the Veil Nebula used in this work. The full list of contributors is given in Table~B.1. DMD acknowledges financial support for a visiting researcher stay at the Astronomy and Astrophysics Department of the University of Valencia within the framework of the «Talent Attraction» programme implemented by the Office of the Vice-Principal for Research (INV25-01-15). 
MPT acknowledges financial support from the Severo Ochoa grant
CEX2021-001131-S and from the Spanish grant PID2023-147883NB-C21, funded by MCIU/AEI/ 10.13039/501100011033, as well as support through ERDF/EU.

The Wendelstein 2.1\,m telescope project was funded by the Bavarian government and by the German Federal government through a common funding process. Part of the 2.1\,m instrumentation, including some of the upgrades for the infrastructure, was funded by the Cluster of Excellence “Origin of the Universe” of the German Science Foundation DFG.\\
This work has made use of data from the European Space Agency (ESA) mission Gaia (https://www.cosmos.esa.int/gaia), processed by the Gaia Data Processing and Analysis Consortium (DPAC, https://www.cosmos.esa.int/web/gaia/dpac/consortium). Funding for the DPAC has been provided by national institutions, in particular the institutions participating in the Gaia Multilateral Agreement.
\end{acknowledgements}

\bibliographystyle{aa} 
\bibliography{Bibliography}

\begin{thebibliography}{33}
\expandafter\ifx\csname natexlab\endcsname\relax\def\natexlab#1{#1}\fi

\bibitem[{Blair {et~al.}(2005)Blair, Sankrit, \& Raymond}]{Blair2005}
Blair, W.~P., Sankrit, R., \& Raymond, J.~C. 2005, The Astronomical Journal, 129, 2268

\bibitem[{{Brownsberger} \& {Romani}(2014)}]{Brownsberger2014}
{Brownsberger}, S. \& {Romani}, R.~W. 2014, ApJ, 784, 154

\bibitem[{{Cant{\'o}} {et~al.}(1998){Cant{\'o}}, {Raga}, {Steffen}, \& {Shapiro}}]{Canto1998}
{Cant{\'o}}, J., {Raga}, A., {Steffen}, W., \& {Shapiro}, P.~R. 1998, \apj, 502, 695

\bibitem[{Cant{\'o} \& Raga(1998)}]{CantoRaga1998}
Cant{\'o}, J. \& Raga, A.~C. 1998, Monthly Notices of the Royal Astronomical Society, 297, 383

\bibitem[{{Chatterjee} \& {Cordes}(2002)}]{Chatterjee2002}
{Chatterjee}, S. \& {Cordes}, J.~M. 2002, ApJ, 575, 407

\bibitem[{{Chatterjee} \& {Cordes}(2004)}]{2004ApJ...600L..51C}
{Chatterjee}, S. \& {Cordes}, J.~M. 2004, \apjl, 600, L51

\bibitem[{{{de Vries}} \& {Romani}(2020)}]{deVries2020}
{{de Vries}}, M. \& {Romani}, R.~W. 2020, ApJL, 896, L7

\bibitem[{{{de Vries}} \& {Romani}(2022)}]{deVries2022}
{{de Vries}}, M. \& {Romani}, R.~W. 2022, ApJ, 928, 39

\bibitem[{{Fesen} \& {Hurford}(1996)}]{Fesen1996}
{Fesen}, R.~A. \& {Hurford}, A.~P. 1996, \apjs, 106, 563

\bibitem[{{Fesen} {et~al.}(2021){Fesen}, {Weil}, {Cisneros}, {Blair}, \& {Raymond}}]{Fesen2021}
{Fesen}, R.~A., {Weil}, K.~E., {Cisneros}, I., {Blair}, W.~P., \& {Raymond}, J.~C. 2021, \mnras, 507, 244

\bibitem[{Fesen {et~al.}(2021)Fesen, Weil, Cisneros, Blair, {et~al.}}]{Fesen2021CygnusLoop}
Fesen, R.~A., Weil, K.~E., Cisneros, I., Blair, W.~P., {et~al.} 2021, Monthly Notices of the Royal Astronomical Society, 507, 261

\bibitem[{{Gaia Collaboration} {et~al.}(2023){Gaia Collaboration}, {Vallenari, A.}, {Brown, A. G. A.}, {Prusti, T.}, {de Bruijne, J. H. J.}, {Arenou, F.}, {Babusiaux, C.}, {Biermann, M.}, {Creevey, O. L.}, {Ducourant, C.}, {Evans, D. W.}, {Eyer, L.}, {Guerra, R.}, {Hutton, A.}, {Jordi, C.}, {Klioner, S. A.}, {Lammers, U. L.}, {Lindegren, L.}, {Luri, X.}, {Mignard, F.}, {Panem, C.}, {Pourbaix, D.}, {Randich, S.}, {Sartoretti, P.}, {Soubiran, C.}, {Tanga, P.}, {Walton, N. A.}, {Bailer-Jones, C. A. L.}, {Bastian, U.}, {Drimmel, R.}, {Jansen, F.}, {Katz, D.}, {Lattanzi, M. G.}, {van Leeuwen, F.}, {Bakker, J.}, {Cacciari, C.}, {Casta\~neda, J.}, {De Angeli, F.}, {Fabricius, C.}, {Fouesneau, M.}, {Fr\'emat, Y.}, {Galluccio, L.}, {Guerrier, A.}, {Heiter, U.}, {Masana, E.}, {Messineo, R.}, {Mowlavi, N.}, {Nicolas, C.}, {Nienartowicz, K.}, {Pailler, F.}, {Panuzzo, P.}, {Riclet, F.}, {Roux, W.}, {Seabroke, G. M.}, {Sordo, R.}, {Th\'evenin, F.}, {Gracia-Abril, G.}, {Portell, J.}, {Teyssier, D.}, {Altmann, M.}, {Andrae,
  R.}, {Audard, M.}, {Bellas-Velidis, I.}, {Benson, K.}, {Berthier, J.}, {Blomme, R.}, {Burgess, P. W.}, {Busonero, D.}, {Busso, G.}, {C\'anovas, H.}, {Carry, B.}, {Cellino, A.}, {Cheek, N.}, {Clementini, G.}, {Damerdji, Y.}, {Davidson, M.}, {de Teodoro, P.}, {Nu\~nez Campos, M.}, {Delchambre, L.}, {Dell\'{}Oro, A.}, {Esquej, P.}, {Fern\'andez-Hern\'andez, J.}, {Fraile, E.}, {Garabato, D.}, {Garc\'{\i}a-Lario, P.}, {Gosset, E.}, {Haigron, R.}, {Halbwachs, J.-L.}, {Hambly, N. C.}, {Harrison, D. L.}, {Hern\'andez, J.}, {Hestroffer, D.}, {Hodgkin, S. T.}, {Holl, B.}, {Jan\ss{}en, K.}, {Jevardat de Fombelle, G.}, {Jordan, S.}, {Krone-Martins, A.}, {Lanzafame, A. C.}, {L\"offler, W.}, {Marchal, O.}, {Marrese, P. M.}, {Moitinho, A.}, {Muinonen, K.}, {Osborne, P.}, {Pancino, E.}, {Pauwels, T.}, {Recio-Blanco, A.}, {Reyl\'e, C.}, {Riello, M.}, {Rimoldini, L.}, {Roegiers, T.}, {Rybizki, J.}, {Sarro, L. M.}, {Siopis, C.}, {Smith, M.}, {Sozzetti, A.}, {Utrilla, E.}, {van Leeuwen, M.}, {Abbas, U.}, {\'Abrah\'am, P.},
  {Abreu Aramburu, A.}, {Aerts, C.}, {Aguado, J. J.}, {Ajaj, M.}, {Aldea-Montero, F.}, {Altavilla, G.}, {\'Alvarez, M. A.}, {Alves, J.}, {Anders, F.}, {Anderson, R. I.}, {Anglada Varela, E.}, {Antoja, T.}, {Baines, D.}, {Baker, S. G.}, {Balaguer-N\'u\~nez, L.}, {Balbinot, E.}, {Balog, Z.}, {Barache, C.}, {Barbato, D.}, {Barros, M.}, {Barstow, M. A.}, {Bartolom\'e, S.}, {Bassilana, J.-L.}, {Bauchet, N.}, {Becciani, U.}, {Bellazzini, M.}, {Berihuete, A.}, {Bernet, M.}, {Bertone, S.}, {Bianchi, L.}, {Binnenfeld, A.}, {Blanco-Cuaresma, S.}, {Blazere, A.}, {Boch, T.}, {Bombrun, A.}, {Bossini, D.}, {Bouquillon, S.}, {Bragaglia, A.}, {Bramante, L.}, {Breedt, E.}, {Bressan, A.}, {Brouillet, N.}, {Brugaletta, E.}, {Bucciarelli, B.}, {Burlacu, A.}, {Butkevich, A. G.}, {Buzzi, R.}, {Caffau, E.}, {Cancelliere, R.}, {Cantat-Gaudin, T.}, {Carballo, R.}, {Carlucci, T.}, {Carnerero, M. I.}, {Carrasco, J. M.}, {Casamiquela, L.}, {Castellani, M.}, {Castro-Ginard, A.}, {Chaoul, L.}, {Charlot, P.}, {Chemin, L.}, {Chiaramida,
  V.}, {Chiavassa, A.}, {Chornay, N.}, {Comoretto, G.}, {Contursi, G.}, {Cooper, W. J.}, {Cornez, T.}, {Cowell, S.}, {Crifo, F.}, {Cropper, M.}, {Crosta, M.}, {Crowley, C.}, {Dafonte, C.}, {Dapergolas, A.}, {David, M.}, {David, P.}, {de Laverny, P.}, {De Luise, F.}, {De March, R.}, {De Ridder, J.}, {de Souza, R.}, {de Torres, A.}, {del Peloso, E. F.}, {del Pozo, E.}, {Delbo, M.}, {Delgado, A.}, {Delisle, J.-B.}, {Demouchy, C.}, {Dharmawardena, T. E.}, {Di Matteo, P.}, {Diakite, S.}, {Diener, C.}, {Distefano, E.}, {Dolding, C.}, {Edvardsson, B.}, {Enke, H.}, {Fabre, C.}, {Fabrizio, M.}, {Faigler, S.}, {Fedorets, G.}, {Fernique, P.}, {Fienga, A.}, {Figueras, F.}, {Fournier, Y.}, {Fouron, C.}, {Fragkoudi, F.}, {Gai, M.}, {Garcia-Gutierrez, A.}, {Garcia-Reinaldos, M.}, {Garc\'{\i}a-Torres, M.}, {Garofalo, A.}, {Gavel, A.}, {Gavras, P.}, {Gerlach, E.}, {Geyer, R.}, {Giacobbe, P.}, {Gilmore, G.}, {Girona, S.}, {Giuffrida, G.}, {Gomel, R.}, {Gomez, A.}, {Gonz\'alez-N\'u\~nez, J.}, {Gonz\'alez-Santamar\'{\i}a, I.},
  {Gonz\'alez-Vidal, J. J.}, {Granvik, M.}, {Guillout, P.}, {Guiraud, J.}, {Guti\'errez-S\'anchez, R.}, {Guy, L. P.}, {Hatzidimitriou, D.}, {Hauser, M.}, {Haywood, M.}, {Helmer, A.}, {Helmi, A.}, {Sarmiento, M. H.}, {Hidalgo, S. L.}, {Hilger, T.}, {Hladczuk, N.}, {Hobbs, D.}, {Holland, G.}, {Huckle, H. E.}, {Jardine, K.}, {Jasniewicz, G.}, {Jean-Antoine Piccolo, A.}, {Jim\'enez-Arranz, \'O.}, {Jorissen, A.}, {Juaristi Campillo, J.}, {Julbe, F.}, {Karbevska, L.}, {Kervella, P.}, {Khanna, S.}, {Kontizas, M.}, {Kordopatis, G.}, {Korn, A. J.}, {K\'osp\'al, \'A}, {Kostrzewa-Rutkowska, Z.}, {Kruszy\'{}nska, K.}, {Kun, M.}, {Laizeau, P.}, {Lambert, S.}, {Lanza, A. F.}, {Lasne, Y.}, {Le Campion, J.-F.}, {Lebreton, Y.}, {Lebzelter, T.}, {Leccia, S.}, {Leclerc, N.}, {Lecoeur-Taibi, I.}, {Liao, S.}, {Licata, E. L.}, {Lindstr\o{}m, H. E. P.}, {Lister, T. A.}, {Livanou, E.}, {Lobel, A.}, {Lorca, A.}, {Loup, C.}, {Madrero Pardo, P.}, {Magdaleno Romeo, A.}, {Managau, S.}, {Mann, R. G.}, {Manteiga, M.}, {Marchant, J. M.},
  {Marconi, M.}, {Marcos, J.}, {Marcos Santos, M. M. S.}, {Mar\'{\i}n Pina, D.}, {Marinoni, S.}, {Marocco, F.}, {Marshall, D. J.}, {Martin Polo, L.}, {Mart\'{\i}n-Fleitas, J. M.}, {Marton, G.}, {Mary, N.}, {Masip, A.}, {Massari, D.}, {Mastrobuono-Battisti, A.}, {Mazeh, T.}, {McMillan, P. J.}, {Messina, S.}, {Michalik, D.}, {Millar, N. R.}, {Mints, A.}, {Molina, D.}, {Molinaro, R.}, {Moln\'ar, L.}, {Monari, G.}, {Mongui\'o, M.}, {Montegriffo, P.}, {Montero, A.}, {Mor, R.}, {Mora, A.}, {Morbidelli, R.}, {Morel, T.}, {Morris, D.}, {Muraveva, T.}, {Murphy, C. P.}, {Musella, I.}, {Nagy, Z.}, {Noval, L.}, {Oca\~na, F.}, {Ogden, A.}, {Ordenovic, C.}, {Osinde, J. O.}, {Pagani, C.}, {Pagano, I.}, {Palaversa, L.}, {Palicio, P. A.}, {Pallas-Quintela, L.}, {Panahi, A.}, {Payne-Wardenaar, S.}, {Pe\~nalosa Esteller, X.}, {Penttil\"a, A.}, {Pichon, B.}, {Piersimoni, A. M.}, {Pineau, F.-X.}, {Plachy, E.}, {Plum, G.}, {Poggio, E.}, {Prsa, A.}, {Pulone, L.}, {Racero, E.}, {Ragaini, S.}, {Rainer, M.}, {Raiteri, C. M.},
  {Rambaux, N.}, {Ramos, P.}, {Ramos-Lerate, M.}, {Re Fiorentin, P.}, {Regibo, S.}, {Richards, P. J.}, {Rios Diaz, C.}, {Ripepi, V.}, {Riva, A.}, {Rix, H.-W.}, {Rixon, G.}, {Robichon, N.}, {Robin, A. C.}, {Robin, C.}, {Roelens, M.}, {Rogues, H. R. O.}, {Rohrbasser, L.}, {Romero-G\'omez, M.}, {Rowell, N.}, {Royer, F.}, {Ruz Mieres, D.}, {Rybicki, K. A.}, {Sadowski, G.}, {S\'aez N\'u\~nez, A.}, {Sagrist\`a Sell\'es, A.}, {Sahlmann, J.}, {Salguero, E.}, {Samaras, N.}, {Sanchez Gimenez, V.}, {Sanna, N.}, {Santove\~na, R.}, {Sarasso, M.}, {Schultheis, M.}, {Sciacca, E.}, {Segol, M.}, {Segovia, J. C.}, {S\'egransan, D.}, {Semeux, D.}, {Shahaf, S.}, {Siddiqui, H. I.}, {Siebert, A.}, {Siltala, L.}, {Silvelo, A.}, {Slezak, E.}, {Slezak, I.}, {Smart, R. L.}, {Snaith, O. N.}, {Solano, E.}, {Solitro, F.}, {Souami, D.}, {Souchay, J.}, {Spagna, A.}, {Spina, L.}, {Spoto, F.}, {Steele, I. A.}, {Steidelm\"uller, H.}, {Stephenson, C. A.}, {S\"uveges, M.}, {Surdej, J.}, {Szabados, L.}, {Szegedi-Elek, E.}, {Taris, F.}, {Taylor,
  M. B.}, {Teixeira, R.}, {Tolomei, L.}, {Tonello, N.}, {Torra, F.}, {Torra, J.}, {Torralba Elipe, G.}, {Trabucchi, M.}, {Tsounis, A. T.}, {Turon, C.}, {Ulla, A.}, {Unger, N.}, {Vaillant, M. V.}, {van Dillen, E.}, {van Reeven, W.}, {Vanel, O.}, {Vecchiato, A.}, {Viala, Y.}, {Vicente, D.}, {Voutsinas, S.}, {Weiler, M.}, {Wevers, T.}, {Wyrzykowski, L.}, {Yoldas, A.}, {Yvard, P.}, {Zhao, H.}, {Zorec, J.}, {Zucker, S.}, \& {Zwitter, T.}}]{gaiadr3}
{Gaia Collaboration}, {Vallenari, A.}, {Brown, A. G. A.}, {et~al.} 2023, A\&A, 674, A1

\bibitem[{{Ghavamian} {et~al.}(2001){Ghavamian}, {Raymond}, {Smith}, \& {Hartigan}}]{Ghavamian2001}
{Ghavamian}, P., {Raymond}, J., {Smith}, R.~C., \& {Hartigan}, P. 2001, The Astrophysical Journal, 547, 995

\bibitem[{Hartigan {et~al.}(2005)Hartigan, Heathcote, Morse, Reipurth, \& Bally}]{Hartigan2005}
Hartigan, P., Heathcote, S., Morse, J., Reipurth, B., \& Bally, J. 2005, Astronomical Journal, 130, 2197

\bibitem[{{Heng}(2010)}]{Heng2010}
{Heng}, K. 2010, Publications of the Astronomical Society of Australia, 27, 23

\bibitem[{{Hester} {et~al.}(1994){Hester}, {Raymond}, \& {Blair}}]{Hester1994}
{Hester}, J.~J., {Raymond}, J.~C., \& {Blair}, W.~P. 1994, The Astrophysical Journal, 420, 721

\bibitem[{{Hester} {et~al.}(1986){Hester}, {Raymond}, \& {Danielson}}]{Hester1986}
{Hester}, J.~J., {Raymond}, J.~C., \& {Danielson}, G.~E. 1986, The Astrophysical Journal Letters, 303, L17

\bibitem[{{Meyer} {et~al.}(2016){Meyer}, {van Marle}, {Kuiper}, \& {Kley}}]{Meyer2016}
{Meyer}, D.~M.-A., {van Marle}, A.-J., {Kuiper}, R., \& {Kley}, W. 2016, MNRAS, 459, 1146

\bibitem[{Morlino {et~al.}(2012)Morlino, Bandiera, Blasi, \& Amato}]{Morlino_2012}
Morlino, G., Bandiera, R., Blasi, P., \& Amato, E. 2012, The Astrophysical Journal, 760, 137

\bibitem[{{Peri} {et~al.}(2012){Peri}, {Benaglia}, {Brookes}, {Stevens}, \& {Isequilla}}]{Peri2012}
{Peri}, C.~S., {Benaglia}, P., {Brookes}, D.~P., {Stevens}, I.~R., \& {Isequilla}, N. 2012, A\&A, 538, A108

\bibitem[{{Peri} {et~al.}(2015){Peri}, {Benaglia}, \& {Isequilla}}]{Peri2015}
{Peri}, C.~S., {Benaglia}, P., \& {Isequilla}, N.~L. 2015, A\&A, 578, A45

\bibitem[{Pippert {et~al.}(2025)Pippert, Kluge, \& Bender}]{Pippert_2025}
Pippert, J.-N., Kluge, M., \& Bender, R. 2025, The Astrophysical Journal, 980, 244

\bibitem[{{Raga} {et~al.}(2000){Raga}, {L{\'o}pez-Mart{\'\i}n}, {Binette}, {L{\'o}pez}, {Cant{\'o}}, {Arthur}, {Mellema}, {Steffen}, \& {Ferruit}}]{Raga2000}
{Raga}, A., {L{\'o}pez-Mart{\'\i}n}, L., {Binette}, L., {et~al.} 2000, \mnras, 314, 681

\bibitem[{{Raga} {et~al.}(2022){Raga}, {Cant{\'o}}, \& {Noriega-Crespo}}]{Raga2022}
{Raga}, A.~C., {Cant{\'o}}, J., \& {Noriega-Crespo}, A. 2022, Revista Mexicana de Astronom{\'i}a y Astrof{\'i}sica, 58, 395

\bibitem[{{Raymond} {et~al.}(2023){Raymond}, {Chilingarian}, {Blair}, {et~al.}}]{Raymond2023}
{Raymond}, J.~C., {Chilingarian}, I.~V., {Blair}, W.~P., {et~al.} 2023, The Astrophysical Journal, 954, 34

\bibitem[{{Reipurth} \& {Bally}(2001)}]{Reipurth2001}
{Reipurth}, B. \& {Bally}, J. 2001, Annual Review of Astronomy and Astrophysics, 39, 403

\bibitem[{{Rom\'an} {et~al.}(2020){Rom\'an}, {Trujillo}, \& {Montes, Mireia}}]{roman+2020}
{Rom\'an}, J., {Trujillo}, I., \& {Montes, Mireia}. 2020, A\&A, 644, A42

\bibitem[{{Sollerman} {et~al.}(2003){Sollerman}, {Ghavamian}, {Lundqvist}, \& {Smith}}]{Sollerman2003}
{Sollerman}, J., {Ghavamian}, P., {Lundqvist}, P., \& {Smith}, R.~C. 2003, Astronomy \& Astrophysics, 407, 249

\bibitem[{{Stupar} {et~al.}(2012){Stupar}, {Parker}, \& {Filipovi{\'c}}}]{Stupar2012}
{Stupar}, M., {Parker}, Q.~A., \& {Filipovi{\'c}}, M.~D. 2012, Monthly Notices of the Royal Astronomical Society, 419, 1413

\bibitem[{{Tetzlaff} {et~al.}(2011){Tetzlaff}, {Neuh{\"a}user}, \& {Hohle}}]{2011MNRAS.410..190T}
{Tetzlaff}, N., {Neuh{\"a}user}, R., \& {Hohle}, M.~M. 2011, \mnras, 410, 190

\bibitem[{Uro{\v{s}}evi{\'c} {et~al.}(2026)Uro{\v{s}}evi{\'c}, Andjeli{\'c}, Filipovi{\'c}, Smeaton, Crawford, Raymond, \& Oni{\'c}}]{Urosevic2026}
Uro{\v{s}}evi{\'c}, D., Andjeli{\'c}, M., Filipovi{\'c}, M.~D., {et~al.} 2026, The radio emission from radiative filaments of Cygnus Loop

\bibitem[{{Vu{\v{c}}eti{\'c}} {et~al.}(2023){Vu{\v{c}}eti{\'c}}, {Milanovi{\'c}}, {Uro{\v{s}}evi{\'c}}, {Raymond}, {Oni{\'c}}, {Milo{\v{s}}evi{\'c}}, \& {Petrov}}]{Vucetic2023}
{Vu{\v{c}}eti{\'c}}, M., {Milanovi{\'c}}, N., {Uro{\v{s}}evi{\'c}}, D., {et~al.} 2023, Serbian Astronomical Journal, 207, 9

\bibitem[{{Wilkin}(1996)}]{Wilkin1996}
{Wilkin}, F.~P. 1996, ApJ, 459, L31

\end{thebibliography}

\begin{appendix}

\section{Determination of the Narrowband Zero Points}\label{app:zeropoints}

\begin{figure}
    \begin{center}
    \includegraphics[width=0.5\textwidth]{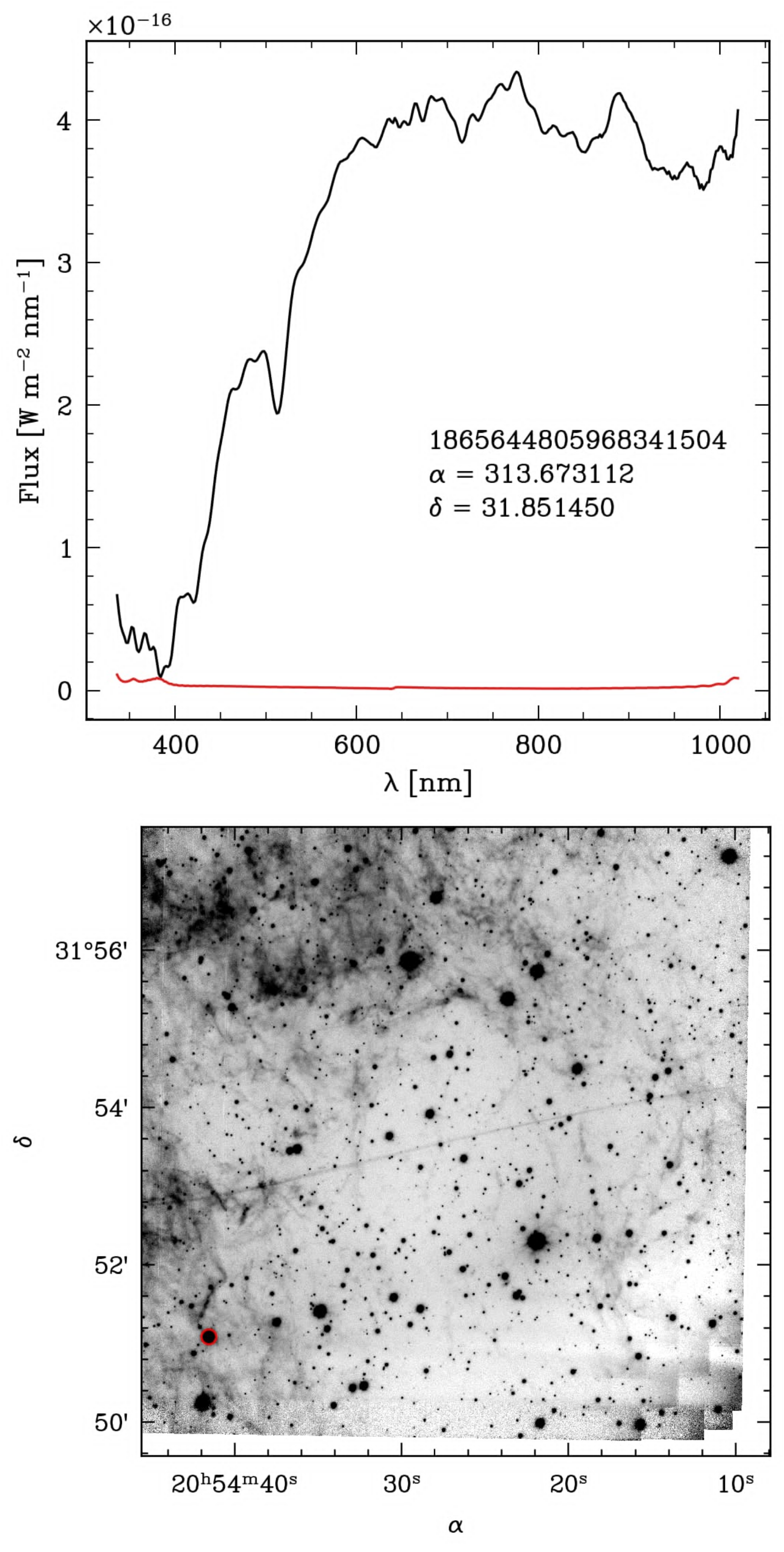}
    \caption{Calibration example of a narrow-band image using Gaia spectra. The red circle in the bottom image highlights the used star from which the spectra (black) and error spectrum (red) are shown in the top panel. The corresponding Gaia ID is shown in the legend.}
    \label{fig:calib}
    \end{center}
\end{figure}

While there is no calibrated narrowband coverage of any survey in that area, we use GaiaDR3 \citep{gaiadr3} low-resolution BP/RP spectra (see Fig.\ref{fig:calib}). From the catalog, 44 sources fall within the 3kk field of view, from which 8 are de-blended or contaminated and are therefore discarded for the zero-point calibration. For each star, aperture photometry is done on the final stacks using \texttt{photutils}, which includes a circular aperture to measure the total flux and a circular annulus to determine the median local background. We estimate the background uncertainty from the standard deviation of the pixels inside the annulus, which can be rather high, as the calibration stars lie within the remnant. For that purpose,  we integrate the corresponding spectrum inside the transmission curves using Equation \ref{eq:flux_int}. The flux has now units of W m$^{-2}$. To transform it into units of W m$^{-2}$ Hz$^{-1}$, hence $F_\nu$, we multiply it by $\frac{\lambda_e^2}{c}$ where $\lambda_e$ is, for narrowband filters, the central wavelength of the transmission curve, and $c$ the speed of light in nm s$^{-1}$. Lastly, we compute the spectral magnitudes using Equation \ref{eq:spec_mag}, where $F_\nu$ is transformed into Jy.

\begin{equation}\label{eq:flux_int}
    F_\lambda = \frac{\int F_\lambda(\lambda)T(\lambda)d\lambda}{\int T(\lambda)d\lambda}
\end{equation}

\begin{equation}\label{eq:spec_mag}
    m_\mathrm{spec} = -2.5\log{\frac{F_\nu}{10^{-26}\times3631}}
\end{equation}

Finally, we compute the difference of both magnitudes for each star and take the median and standard deviation to arrive at our final photometric zero-points (Table \ref{tab:zeropoints}) for the $H_\alpha$ and $SII$ filters.

\begin{table}[h]
    \centering
    \begin{tabular}{c|c|c}
     Filter &  Exposure time [s] & Zeropoint [mag]  \\
        \hline
        \hline
        $H_\alpha$ & 14880 & $21.05\pm0.13$ \\
        $SII$      & 11880 & $21.15\pm0.11$ \\
        $i$        & 26460 & $30.38$\\ 
    \end{tabular}
    \caption{Total exposure time and zero points for the Wendelstein CCD 3kk stack images. The $i$-band zeropoint was directly determined by the 3kk data reduction pipeline}
    \label{tab:zeropoints}
\end{table}

\section{Detection of the trail feature in amateur images}
\label{appendix:b}

\begin{figure*}
    \begin{center}
    \includegraphics[width=0.9\textwidth]{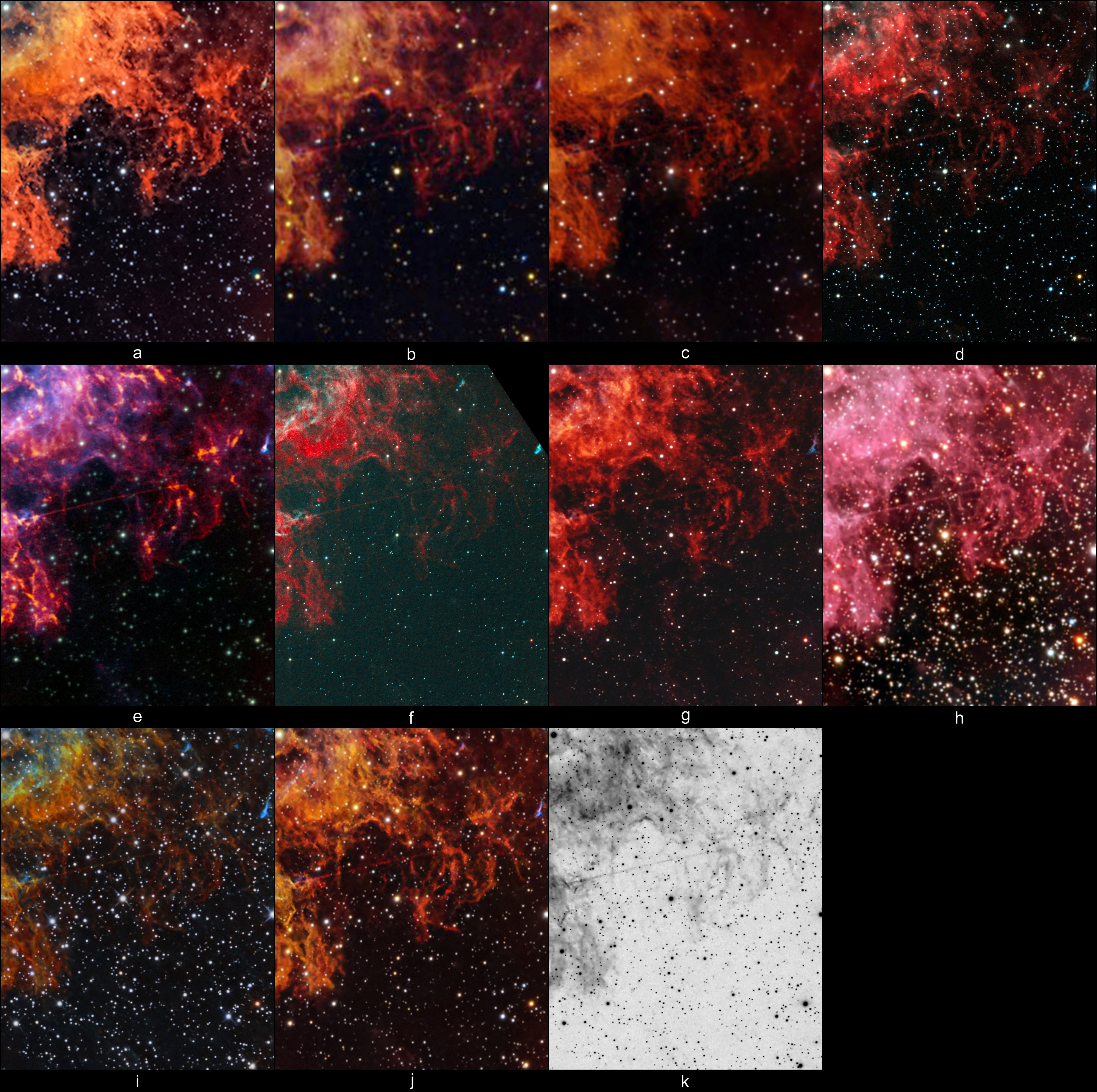}
    \caption{ Multi-epoch images of the Veil Nebula trail feature obtained with different instruments. No significant displacement or morphological changes of the trail is detected. Details of these observations are listed in Table B.1.}
    \label{fig:amateurs}
    \end{center}
\end{figure*}

The morphology of the trail-like feature is reminiscent of signatures produced by artificial satellites in astronomical images, suggesting that it could be an observational artifact rather than a physical structure associated with the Veil Nebula. We therefore have examined independent observations of the Veil Nebula obtained at different epochs using a variety of amateur equipment distributed among the astrophotography community. We have compiled a set of publicly available amateur images spanning several years (see Table B.1), all of which clearly show this structure (Fig.\ref{fig:amateurs}). This completely rejects the possibility of an artificial origin (e.g., satellite trails or other acquisition artifacts) and confirms it as a real structure associated with (or projected onto) the Veil Nebula.

\begin{table*}
    \caption{Confirmation images of the ´satellite train´ in  the SNR Veil}
    
    \begin{tabular}{l|l|c|c|l|c|l}
     Telescope & Detector & Date & Image Scale & Astrophotographer & Figure  \\
        \hline
 Askar FRA400 f/2.5  & ZWO ASI2600MM Pro & Jul 2024 & 2.53 & Victor(Astro) BG  & Fig. 7a  \\
    Takahashi FSQ-106EDX III f/5 & MI G4-16000 MK & Jul-Nov 2024 &
    3.26 & Alpha Zhang & Fig. 7b \\
    William Optics Pleiades 68 f/2.6 & ZWO ASI6200MM Pro & May-Jun 2025 & 2.58 & Jeko Jekov & Fig. 7c \\
    Perl Bellatrix 200/1000 f/5.0 & ZWO ASI2600MC DUO & Jul 2025 & 1.02 & Cl\`ement Daniel & Fig. 7d \\
    Takahashi FSQ-106EDX4 f/5.0 & ZWO ASI6200MM Pro & Jul 2025 & 2.01 & Bray Falls &  Fig. 7e \\
    Sky-Watcher Quattro 250P/10-S f/4.0 & ZWO ASI2600MC Pro &  Aug 2025 & 0.77 & Jan Hlavacka & Fig 7f \\
    Celestron C9.25 SC XLT f/10 & ZWO ASI533MC Pro & Sep  2025 & 1.51 & David Foust & Fig. 7g \\
    TS-Optics 86SDQ APO f/5.4 & ZWO ASI6200MM Pro & Sep-Dec 2025 &  2.14 & Brent Newton & Fig. 7h \\
    Takahashi Epsilon-160ED f/3.3  & MI C1 61000 Pro & Jul 2025 & 1.45 & Kenward Vaughan & Fig. 7i \\
    William Optics 91/FLT91 f/5.9 & ZWO ASI6200MM Pro & Jul-Aug 2025 & 1.78 & Rafa\l{} Szwejkowski & Fig. 7j \\
    Takahashi FSQ106EDX f/5  & Atik 16200 Mono & Aug 2024 & 2.34 & \'Alvaro Ib\'a\~nez P\'erez & Fig. 7k \\

\hline
\end{tabular}
\end{table*}
 
\end{appendix}
\end{document}